\documentclass[a4paper,cits]{PoS}

\title{Exploring confinement in SU(N) gauge theories with double-trace Polyakov loop deformations}

\ShortTitle{Exploring confinement in SU(N) gauge theories with double-trace Polyakov loop deformations}

\author{\speaker{Michael Ogilvie}\\
       Washington University, St. Louis\\
       E-mail: \email{mco@physics.wustl.edu}}

\author{Peter Meisinger\\
        Washington University, St. Louis\\
        E-mail: \email{pnm@physics.wustl.edu}}

\abstract{Recent results applying resurgence theory to finite-temperature field
theories yield a detailed analytic structure determined by topological
excitations. We examine finite-temperature SU(N) lattice gauge theories
in light of these results. Double-trace Polyakov loop deformations
move through different regions of the confined phase characterized
by continuous change in the adjoint Polyakov loop. Lattice models
show how the behavior of monopole constituents of calorons can change
in the different confining regions. We conjecture that the pure SU(N)
gauge theory is close to a special symmetric point where monopole
effects give rise to Casimir string-tension scaling.
}

\FullConference{The 32nd International Symposium on Lattice Field Theory\\
		 23-28 June, 2014\\
		 Columbia University New York, NY}

\begin{document}

\section{Introduction}

Topological features of four-dimensional
gauge theories on $R^{3}\times S^{1}$, deformed from a pure gauge theory by either
a double-trace deformation or by periodic adjoint fermions, have been
been shown to lead to confinement in a region where semiclassical methods are
valid \cite{Myers:2007vc,Unsal:2007vu,Unsal:2007jx}. Moreover, this
region is smoothly connected to the usual low-temperature confining
region \cite{Myers:2007vc}. Recents results applying resurgence theory
to such models \cite{Argyres:2012vv,Argyres:2012ka} suggest
a detailed analytic structure determined by the interplay of perturbative
and non-perturbative physics. The behavior of observables are described
by a trans-series
\begin{equation}
\left\langle O\right\rangle =\sum_{n=0}^{\infty}p_{0,n}\lambda^{n}+\sum_{c}e^{-S_{c}/\lambda}\sum_{n=0}^{\infty}p_{c,n}\lambda^{n}.
\end{equation}
This sum includes contributions which are not topologically stable,
such as an instanton-anti-instanton contributions, and applies even
in theories without topologically stable contributions \cite{Cherman:2013yfa}.
In light of these results, we want to explore the structure of the
confining phase under the action of double-trace Polyakov loop deformations.
Such deformations interpolate between different Abelian
limiting models and change the interpretation of topological excitations
and their role. In lattice field theories, duality can be used in
place of semiclassical continuum techniques to obtain the topological
content of deformed field theories. What emerges from the interplay
between continuum and lattice models is a consistent picture of confinement
in which the non-universal weighting of topological objects determines
string tension scaling.

\section{The $O(3)$ model in $d=2$}

A simple model of the behavior we wish to study in gauge theories
is afforded by the $O(3)$ model in $d=2$ space-time dimensions.
The continuum Euclidean action is given by
\begin{equation}
S=\int d^{2}x\frac{1}{2g^{2}}\left(\nabla\vec{\sigma}\right)^{2}
\end{equation}
 where the fields are constrained to
 $\vec{\sigma}^{2}=\sigma_{1}^{2}+\sigma_{2}^{2}+\sigma_{3}^{2}=1$.
Like QCD, the $O(3)$ model is asymptotically free and has instantons.

We can deform the $O(3)$ model into an $XY$-like model by deforming
the action \cite{Ogilvie:1981yw,Affleck:1985jy}
\begin{equation}
S\rightarrow S-\int d^{2}x\,\frac{1}{2}h\sigma_{3}^{2}.
\end{equation}
With $h<0$, $\sigma_{3}=0$ is preferred; as $h$ becomes increasingly
negative, the model approaches the $XY$, or $O(2)$ limit. The $XY$
model in $d=2$ has the well-known Kosterlitz-Thouless critical behavior, 
associated with the transition from a low-temperature
($g^{2}$ small) phase dominated by massless spin-waves to a high-temperature
phase ($g^{2}$ large), a vortex-antivortex plasma with a mass gap.
Each Kosterlitz-Thouless vortex-antivortex
pair are the constituents of an $O(3)$ instanton, with
$\sigma_3 \rightarrow \pm 1$ at each vortex or antivortex core.
It is also interesting to consider the behavior of the deformed $O(3)$
model for $h>0$. In this case, the boundary conditions are $\sigma_{3}\rightarrow\pm1$
in the limit $\left|z\right|\rightarrow\infty$, and the system becomes
increasingly like the Ising model as $h\rightarrow\infty$. 
The antivortex moves to infinity
and the interpretation of an instanton
as a vortex-antivortex pair is lost. 
For $h>0$, instantons look like flipped spins in an Ising model low-$T$
expansion: At infinity, $\sigma_{3}\rightarrow-1$ and $\sigma_{3}\rightarrow+1$
at the center of the instanton. 


The deformed model has three distinct phases, as
sketched in Figure \ref{fig:O3-phase-diagram}. There is an Ising-like
phase, where the $Z(2)$ symmetry $\sigma_{3}\rightarrow-\sigma_{3}$
is spontaneously broken. There is a massless spin-wave phase with
the behavior of the low-temperature phase of the XY model. There is
a massive, disordered phase with the $O(3)$ model intermediate
between the XY and Ising limits. The expectation value of $\sigma_{3}^{2}$,
which is conjugate to $h$, determines where we are in the phase diagram.
The Ising limit corresponds to $\left\langle \sigma_{3}^{2}\right\rangle \rightarrow1$
while the XY limit is obtained when $ $$\left\langle \sigma_{3}^{2}\right\rangle \rightarrow0$.
For both $h>0$ and $h<0$, instantons disorder the system, but their
role appears completely different. If we are deep in the $Z(2)$ broken
phase, with $\left\langle \sigma_{3}\right\rangle \approx1$, instantons
are highly suppressed and the dilute instanton gas approximation may
be used. In order to approach the critical line, instantons must lower
$\left\langle \sigma_{3}\right\rangle $ towards zero, and the dilute
instanton gas approximation must fail. In the massless XY phase, vortices
are tightly bound in pairs, and instantons play no role in large-distance
behavior. After crossing the $O(2)$ critical line, a Coulomb gas
of vortices and antivortices forms, leading to a massive phase. It
is only in this region that we have good analytic control of the behavior
of topological excitations outside of the dilute instanton gas approximation.
If we consider $O(N)$ models with $N>3$, it is physically obvious
that they must have Ising and XY limits obtained using a similar deformation
and a similar phase diagram. Although instantons in these higher $O(N)$
models are not topologically stable, the inclusion
of topological excitations is absolutely necessary to reproduce the
phase structure. This is a confirmation of a key assumption of resurgence
theory as applied to quantum field theories: non-topologically stable
solutions matter. 
\begin{figure}
\centerline{\includegraphics[width=4in]{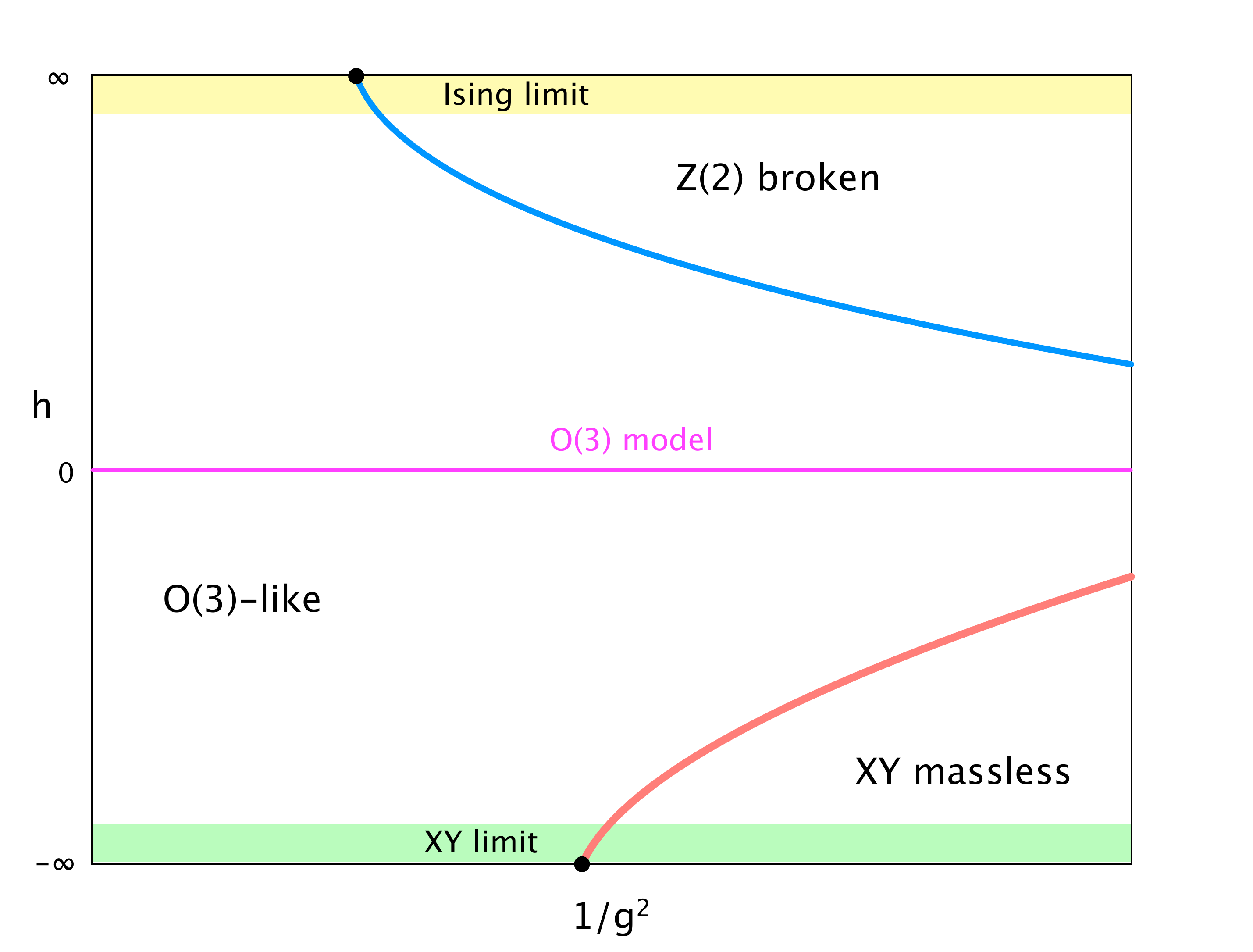}}
\caption{\label{fig:O3_phase_diagram}The phase diagram of the deformed $O(3)$
model as a function of $1/g^{2}$ and $h$.}
\end{figure}

\section{High-T confinement on $R^{3}\times S^{1}$}

Gauge theories on $R^{3}\times S^{1}$ can be treated in a manner
similar to the deformed $O(3)$ model discussed above. It is often
convenient to identify the circumference $L$ of $S^{1}$ with the
inverse temperature $T^{-1}$. In the limit where $T$ is large, two
things occur: The coupling $g^{2}(T)\rightarrow0$ gets weak as $T\rightarrow\infty$,
and global $Z(N)$ is spontaneously broken. It is possible to modify
the action to restore $Z(N)$ symmetry with a so-called double-trace
deformation. This is a term added to the action, depending only on
Polyakov loops in the adjoint representation, that favors the confined,
$Z(N)$-unbroken phase. For the gauge group $SU(2)$, a simple choice
for such a deformation is \cite{Myers:2007vc}
\begin{equation}
S\rightarrow S-\int d^{2}x\, H_{A}\left|Tr_{F}P\right|^{2}
\end{equation}
with $H_{A}<0$. For larger gauge groups, a more complicated deformation
is required \cite{Unsal:2008ch,Myers:2009df,Meisinger:2009ne}. It
is also possible to restore $Z(N)$ symmetry using fermions in the
adjoint representation with periodic boundary conditions \cite{Unsal:2007vu,Unsal:2007jx}.
In the high-$T$ limit, $A_{4}$ behaves as a three-dimensional scalar
with a vacuum expectation value, leading to Euclidean monopole solutions.
These in turn give rise to confinement in ``spatial'' Wilson loops
lying in $R^{3}$ at constant $x_{4}$.

The Euclidean monopoles are constituents of instantons \cite{Lee:1997vp,Lee:1998bb,Kraan:1998kp,Kraan:1998pm}
and confine \cite{Unsal:2007vu,Unsal:2008ch}. Dimensional reduction
yields confinement by monopole gas
as in the $d=3$ Georgi-Glashow model \cite{Polyakov:1976fu}. 
The monopole gas is represented by a sine-Gordon
model for SU(2):
\begin{equation}
S_{eff}=\int d^{3}x\,\left[\frac{g^{2}\left(T\right)T}{32\pi^{2}}\left(\partial_{j}\sigma\right)^{2}-4y\cos\left(\sigma\right)\right]
\end{equation}
where $y\propto T^{3}\left(\Lambda/T\right)^{11/3}$. This program
works in lattice gauge theories lattice as well \cite{Ogilvie:2012fe,Ogilvie:2014bwa}.
In the lattice case, Abelian lattice duality methods can be used to
uncover vortex effects, as opposed to steepest descent methods in
continuum models. On the lattice, vortices of higher charge appear
naturally. They do not have the instabilities seen in continuum analysis
of higher-charge solutions, because on the lattice integrals over
translational zero modes are replaced by lattice sums.

The effect of $H_{A}$ on the gauge theory is similar to that of $h$
on the $O(3)$ model, with some differences. Negative $H_{A}$ increases
the deconfinement temperature, while positive $H_{A}$ promotes $Z(2)$
breaking and decreases the deconfinement temperature. The observable
$Tr_{A}P$ indicates where we are in the phase diagram of the deformed
$SU(2)$ gauge theory. The limit $H_{A}\rightarrow\infty$ is an Ising
limit, where $Tr_{A}P\rightarrow2$ and $Tr_{F}P\rightarrow\pm1$.
When $H_{A}$ is sufficiently negative, we obtain $Tr_{A}P\sim-1$
and $Tr_{F}P\sim0$, and the gauge theory behaves as a $U(1)$ gauge
theory at large distances. The pure gauge theory, with $H_{A}=0$,
has both $Tr_{A}P\sim0$ and $Tr_{F}P\sim0$, and sits between the
two limiting behaviors. A special feature of the deformed $SU(2)$
model is the existence of a tricritical point where the deconfinement
transition changes from 2nd-order to 1st-order. This new critical
point is in the same universality class as the tricritical point in
the Blume-Emery-Griffiths (BEG) model \cite{Blume:1971zza}, and has
non-Ising critical indices. The observable $Tr_{A}P$ plays a role
in the gauge theory analogous to the role of vacancies in the BEG
model.
\begin{figure}
\centerline{\includegraphics[width=4in]{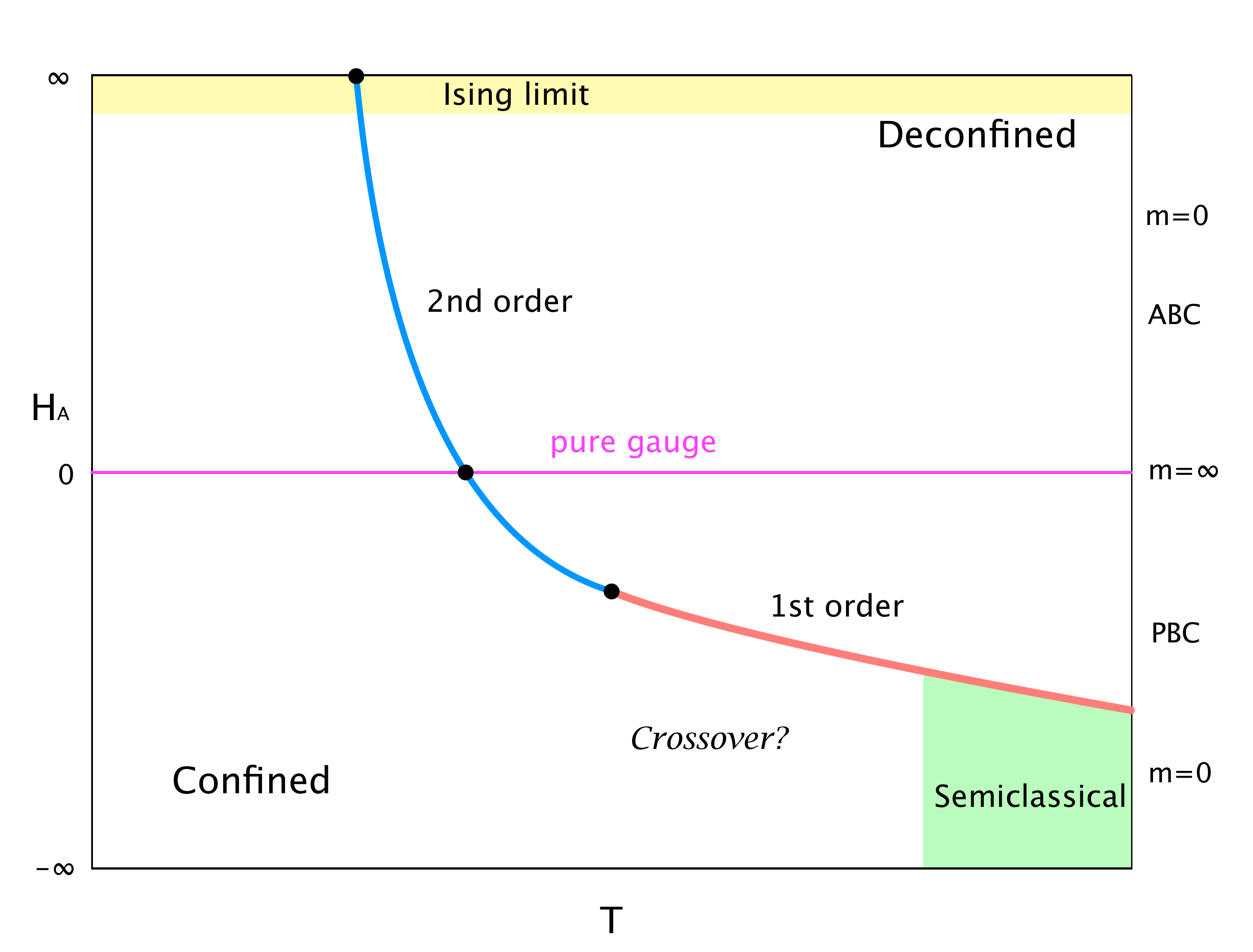}}
\caption{The phase diagram of the deformed $SU(2)$ gauge theory as a function
of $T$ and $H_{A}$.}
\end{figure}

In the semiclassical region of SU(N), inclusion of only the lightest
monopole states gives a generalized sine-Gordon model based on an
$N$-component field $\rho$ with action
\begin{equation}
S_{mag}=\int d^{3}x\,\left[\frac{T}{2}\left(\partial_{j}\rho\right)^{2}-2 y\sum_{k=1}^{N}\cos\left(\frac{2\pi}{g}\alpha_{k}\cdot\rho\right)\right]
\end{equation}
where the $\alpha_{k}$ are the affine roots of SU(N) \cite{Unsal:2008ch}.
This model is one of a whole class of generalized sine-Gordon models
that differ in how different weights are included. In general, string
tensions for different $N$-alities may be calculated as the surface
tension of kink solutions connecting different vacua.
the lowest string tension is known analytically. 
If all roots
are included with equal weighting, an ansatz of straight-line motion
in the Lie algebra leads to Casimir scaling \cite{Giovannangeli:2001bh}.
If only positive roots are included, sine-law scaling
is obtained  \cite{Diakonov:2007nv}.
Thus any deformation that changes the weighting of monopoles
configurations in the partition function may also change string-tension
scaling laws.

\section{Lattice models}

Using the ideas from previous sections, it is now easy to construct
a lattice gauge theory with Casimir scaling. It is an Abelian theory
with a $U(1)^{N-1}$ gauge group and a global $Z(N)$ symmetry in
an $R^{3}\times S^{1}$geometry. We begin with a $U(1)^{N}$ lattice
gauge theory with a Villain action \cite{Banks:1977cc}:
\begin{equation}
S_{1}=\frac{1}{2g^{2}}\sum_{a=1}^{N}\sum_{p}Tr\left(\partial_{\mu}\phi_{\nu}^{a}-\partial_{\nu}\phi_{\mu}^{a}-2\pi n_{\mu\nu}^{a}\right)^{2}
\end{equation}
The integer-valued plaquette variables $n_{\mu\nu}^{a}$ are summed
over all integers to enforce the symmetry. We define a set of monopole
currents:
\begin{equation}
m_{\mu}^{a}\left(X\right)=\frac{1}{2}\epsilon_{\mu\nu\rho\sigma}\partial_{\nu}n_{\rho\sigma}^{a}\left(x\right)
\end{equation}
The remaining degrees of freedom can be integrated out, giving a Coulomb
gas representation:
\begin{equation}
S_{dual}=\frac{2\pi^{2}}{g^{2}}\sum_{a=1}^{N}\sum_{R,R'}m_{\mu}^{a}\left(R\right)G\left(R-R'\right)m_{\mu}^{a}\left(R'\right)
\end{equation}
We can restrict $U(1)^{N}$ to $U(1)^{N-1}$ using a periodic delta
function:
\begin{equation}
S_{2}=-i\sum_{x,\mu}p_{\mu}(x)\left[\sum_{a}\phi_{\mu}^{a}\left(x\right)\right]
\end{equation}
This gives rise to an electric interaction in a Coulomb gas representation
\begin{eqnarray}
S_{dual} & = & \frac{2\pi^{2}}{g^{2}}\sum_{R,R'}m_{\mu}^{a}\left(R\right)G\left(R-R'\right)m_{\mu}^{a}\left(R'\right)+\frac{g^{2}}{2}\sum_{r,r'}p_{\mu}\left(r\right)G\left(r-r'\right)p_{\mu}\left(r'\right)\nonumber \\
 &  & -i\sum_{r,R}\left(\sum_{a}m_{\mu}^{a}\left(R\right)\right)\Theta_{\mu\nu}\left(R-r\right)p_{\nu}\left(r\right)
\end{eqnarray}
representing the interaction of electric and magnetic charges; see,
\emph{e.g.}, \cite{Cardy:1981qy,Cardy:1981fd}. As in continuum field
theories, we can add a term to the action that favors
or disfavors the $Z(N)$ center subgroup of $SU(N)$. On $R^{3}\times S^{1}$,
we can ensure that the dominant terms will be short monopole world lines with $m_{4}^{a}=+1$
and $m_{4}^{b}=-1$ for pairs $\left(a,b\right)$ with $a\ne b$.
Unlike the continuum case, an expectation value for $A_{4}$, and
hence the Polyakov loop, does not play an integral role. All roots
are naturally included with equal weight. After dimensional reduction
and transformation of the monopole gas to a sine-Gordon form, this
leads naturally to Casimir scaling.

\section{Conclusions}

Resurgence theory suggests the need to include a large class of non-perturbative
phenomena in quantum field theories, considerably beyond what has
heretofore been included. Deformations allow us to change the non-perturbative
content in non-trivial ways. In the case of $SU(N)$ gauge theories
on $R^{3}\times S^{1}$, we can interpolate between an $U(1)^{N-1}$
instanton gas picture of confinement and a $Z(N)$ gauge theory, with
the pure gauge theory in the middle, while remaining in the confining
phase. $Tr_{A}P$ indicates where we are in the phase diagram, with
the system behaving as a generalization of the BEG model. Instantons
change their role in moving between regions, but are important throughout.
It seems plausible that these deformations may change string tension
ratios among different $N$-alities. Casimir scaling in particular
is associated with the inclusion of monopoles on a democratic basis,
and appears naturally in a $U(1)^{N-1}$ lattice model.


\begin{thebibliography}{99}

\bibitem{Myers:2007vc} 
  J.~C.~Myers and M.~C.~Ogilvie,
  \emph{New phases of SU(3) and SU(4) at finite temperature},
  \emph{Phys.\ Rev.\ D} {\bf 77}, 125030 (2008)
  [arXiv:0707.1869 [hep-lat]].


\bibitem{Unsal:2007vu} 
  M.~Unsal,
  \emph{Abelian duality, confinement, and chiral symmetry breaking in QCD(adj)},
  \emph{Phys.\ Rev.\ Lett.\ }  {\bf 100}, 032005 (2008)
  [arXiv:0708.1772 [hep-th]].
    
\bibitem{Unsal:2007jx} 
  M.~Unsal,
  \emph{Magnetic bion condensation: A New mechanism of confinement and mass gap in four dimensions},
  \emph{Phys.\ Rev.\ D} {\bf 80}, 065001 (2009)
  [arXiv:0709.3269 [hep-th]].
    
\bibitem{Argyres:2012vv} 
  P.~Argyres and M.~Unsal,
  \emph{A semiclassical realization of infrared renormalons},
  \emph{Phys.\ Rev.\ Lett.\ } {\bf 109}, 121601 (2012)
  [arXiv:1204.1661 [hep-th]].
  
\bibitem{Argyres:2012ka} 
  P.~C.~Argyres and M.~Unsal,
  \emph{The semi-classical expansion and resurgence in gauge theories: new perturbative, instanton, bion, and renormalon effects},
  \emph{JHEP} {\bf 1208}, 063 (2012)
  [arXiv:1206.1890 [hep-th]].
  
  
  
  [arXiv:1306.4405 [hep-th]].
  
\bibitem{Cherman:2013yfa} 
  A.~Cherman, D.~Dorigoni, G.~V.~Dunne and M.~†nsal,
  \emph{Resurgence in Quantum Field Theory: Nonperturbative Effects in the Principal Chiral Model},
  \emph{Phys.\ Rev.\ Lett.\ } {\bf 112}, no. 2, 021601 (2014)
  [arXiv:1308.0127 [hep-th]].
  
\bibitem{Ogilvie:1981yw} 
  M.~C.~Ogilvie and G.~S.~Guralnik,
  \emph{Instantons And Vortices In Two-dimensions},
  \emph{Nucl.\ Phys.\ B} {\bf 190}, 325 (1981).
  
\bibitem{Affleck:1985jy} 
  I.~Affleck,
  \emph{Mass Generation By Merons In Quantum Spin Chains And The O(3) Sigma Model},
  \emph{Phys.\ Rev.\ Lett.\  }{\bf 56}, 408 (1986).
    
  
\bibitem{Unsal:2008ch} 
  M.~Unsal and L.~G.~Yaffe,
  \emph{Center-stabilized Yang-Mills theory: Confinement and large N volume independence},
  \emph{Phys.\ Rev.\ D} {\bf 78}, 065035 (2008)
  [arXiv:0803.0344 [hep-th]].
  
\bibitem{Myers:2009df} 
  J.~C.~Myers and M.~C.~Ogilvie,
  \emph{Phase diagrams of SU(N) gauge theories with fermions in various representations},
  \emph{JHEP} {\bf 0907}, 095 (2009)
  [arXiv:0903.4638 [hep-th]].
 
\bibitem{Meisinger:2009ne} 
  P.~N.~Meisinger and M.~C.~Ogilvie,
  \emph{String Tension Scaling in High-Temperature Confined SU(N) Gauge Theories},
  \emph{Phys.\ Rev.\ D} {\bf 81}, 025012 (2010)
  [arXiv:0905.3577 [hep-lat]].

\bibitem{Lee:1997vp} 
  K.~M.~Lee and P.~Yi,
  \emph{Monopoles and instantons on partially compactified D-branes},
  \emph{Phys.\ Rev.\ D} {\bf 56}, 3711 (1997)
  [hep-th/9702107].

\bibitem{Lee:1998bb} 
  K.~-M.~Lee and C.~-h.~Lu,
  \emph{SU(2) calorons and magnetic monopoles},
  \emph{Phys.\ Rev.\ D} {\bf 58}, 025011 (1998)
  [hep-th/9802108].
      
\bibitem{Kraan:1998kp} 
  T.~C.~Kraan and P.~van Baal,
  \emph{Exact T duality between calorons and Taub - NUT spaces},
  \emph{Phys.\ Lett.\ B} {\bf 428}, 268 (1998)
  [hep-th/9802049].
  
\bibitem{Kraan:1998pm} 
  T.~C.~Kraan and P.~van Baal,
  \emph{Periodic instantons with nontrivial holonomy},
  \emph{Nucl.\ Phys.\ B} {\bf 533}, 627 (1998)
  [hep-th/9805168].
  
\bibitem{Polyakov:1976fu} 
  A.~M.~Polyakov,
  \emph{Quark Confinement and Topology of Gauge Groups},
  \emph{Nucl.\ Phys.\ B} {\bf 120}, 429 (1977).
  
\bibitem{Ogilvie:2012fe} 
  M.~Ogilvie,
  \emph{Confinement in high-temperature lattice gauge theories},
  \emph{PoS LATTICE} {\bf 2012}, 085 (2012)
  [arXiv:1211.1358 [hep-lat]].
  
\bibitem{Ogilvie:2014bwa} 
  M.~C.~Ogilvie,
  \emph{Confinement on $R^{3}\times S^{1}$: continuum and lattice},
  \emph{Int.\ J.\ Mod.\ Phys.\ A} {\bf 29}, 1445003 (2014)
  [arXiv:1410.1860 [hep-th]].
  
\bibitem{Blume:1971zza} 
  M.~Blume, V.~J.~Emery and R.~B.~Griffiths,
  \emph{Ising Model for the lamda Transition and Phase Separation in He-3- He-4 Mixtures},
  \emph{Phys.\ Rev.\ A} {\bf 4}, 1071 (1971).
  
\bibitem{Giovannangeli:2001bh} 
  P.~Giovannangeli and C.~P.~Korthals Altes,
  \emph{'t Hooft and Wilson loop ratios in the QCD plasma},
  \emph{Nucl.\ Phys.\ B} {\bf 608}, 203 (2001)
  [hep-ph/0102022].
  
\bibitem{Diakonov:2007nv} 
  D.~Diakonov and V.~Petrov,
  \emph{Confining ensemble of dyons},
  \emph{Phys.\ Rev.\ D} {\bf 76}, 056001 (2007)
  [arXiv:0704.3181 [hep-th]].
  
  
\bibitem{Banks:1977cc} 
  T.~Banks, R.~Myerson and J.~B.~Kogut,
  \emph{Phase Transitions in Abelian Lattice Gauge Theories},
  Nucl.\ Phys.\ B {\bf 129}, 493 (1977).
 
\bibitem{Cardy:1981qy} 
  J.~L.~Cardy and E.~Rabinovici,
  \emph{Phase Structure of Z(p) Models in the Presence of a Theta Parameter},
  \emph{Nucl.\ Phys.\ B} {\bf 205}, 1 (1982).
  
\bibitem{Cardy:1981fd} 
  J.~L.~Cardy,
  \emph{Duality and the Theta Parameter in Abelian Lattice Models},
  \emph{Nucl.\ Phys.\ B} {\bf 205}, 17 (1982).
      
   

\end{thebibliography}

\end{document}